\documentstyle[12pt]{article}

\def\bq{\begin{quotation}}
\def\eq{\end{quotation}}

\def\fnote#1#2 {\begingroup \def \thefootnote {#1}
\footnote{#2}\addtocounter{footnote}{-1}\endgroup}

\newcommand{\support} {This work is supported in part by U.S. Department of
Energy Contract No.  DE-AC02-76ER13065.}
\newcommand{\myname} {\vspace{0.5in}
                                  \begin{center}Zhu Yang\\
                                  \vspace{0.2in}
                                  Department of Physics and Astronomy\\
                                  University of Rochester\\
                                  Rochester, NY 14627\\
                                  bitnet:yang@uorhep\\
                                 \vspace{0.5in}
                                 Abstract\\
                                  \vspace{0.2in}
                                \end{center}}

\newcommand{\pagenumber}{\pagestyle{plain}\setcounter{page}{1}}

\def\a{\alpha} 
\def\b{\beta} 
 
\def\e{\epsilon}

\def\m{\mu}

\def\raisenot{\raise .5mm\hbox{/}}
\newcommand{\notpa}{\hbox{{$\partial$}\kern-.54em\hbox{\raisenot}}}
\def\notp{\ \hbox{{$p$}\kern-.43em\hbox{/}}}
\def\notq{\ \hbox{{$q$}\kern-.47em\hbox{/}}}
\def\notk{\ \hbox{{$k$}\kern-.47em\hbox{/}}}
\def\notA{\ \hbox{{$A$}\kern-.47em\hbox{/}}}
\def\nota{\ \hbox{{$a$}\kern-.47em\hbox{/}}}
\def\notb{\ \hbox{{$b$}\kern-.47em\hbox{/}}}
\def\notD{\ \hbox{{$D$}\kern-.50em\hbox{/}}} 

\begin {document}
\baselineskip=24pt

\pagestyle{empty}

\begin{flushright}
UR-1264\\
ER-40685-718\\
hep@xxx/9207014\\
\end{flushright}
\vspace{0.5in}
\begin{center}
{\Large A Note on ``Anomalous" Chiral Gauge Theories }
\end{center}
\myname

We show that, the lattice regularization of chiral gauge theories proposed
by Kaplan,  when applied to a (2+1)-dimensional domain wall,
produces a (1+1)-dimensional theory at low energy
even if gauge anomaly produced by
chiral fermions does not cancel. But
the corresponding statement is not true in higher dimensions.

\newpage
\pagenumber
Chiral gauge theories deserve deeper study because nature exhibits left-right
asymmetry.
Recently Kaplan \cite{1} has proposed a new lattice
regularization of chiral gauge
theories (see also \cite{2}). The idea is to use the fermion zero modes trapped
in a 2n+1  dimensional domain wall as matter fields of 2n dimensions.
Under suitable conditions, the fermion is chiral even on a
lattice. The construction
reproduces correct anomalous Ward identities. There is one catch, however:
One has to make sure that the gauge fields live only in 2n dimensions as well.
Kaplan argues that this can be true if and only if gauge currents in
the 2n dimensional theory is anomaly free. Thus the method can be a
very powerful tool to study non-perturbative aspects of chiral gauge
theories. Some numerical results have already been obtained in 1+1
dimension chiral Schwinger model \cite{3}.

On the other hand, 2n dimensional anomalous chiral gauge theories do not
behave in the same way for all $n$. For example, an
anomalous chiral gauge theory can be consistently quantized in D=2
at all energy scales,
but not in D=4 or higher, see \cite{4} for a review. Thus it would
be disturbing if the Kaplan regularization
did not comply with this fact, since it is so promising. In this note
we show that   it does. The difference between D=2 and $D>2$ theories,
in the present language, is due to the role of Chern-Simons terms
in D+1 dimensions. So the Kaplan proposal gives a satisfactory lattice
regularization of ``anomalous" 1+1 dimensional chiral gauge theory.

Since the problem arises in the continuum language as well, we will
not use lattice cutoff explicitly.
Following Kaplan, consider a fermion in 2n+1 dimensions with a position
dependent mass,
\begin{equation}
L= \bar{\psi} (\notpa + m(s))\psi
\end{equation}
where $m(s)$ satisfies  is a step function with $m(0)=0$.
It can be shown that the fermion zero mode is chiral, due to the
normalizability of the zero modes.

When coupled to some gauge field, the massive fermions  produce a
Goldstone-Wilczek\cite{5}  current \cite{6,7,8}
\begin{equation}
J_{\m} \sim
{\rm sgn} m(s) \e^{\a_{1}\b_{1}\dots\a_{n}\b_{n}}F_{\a_{1}\b_{1}}\dots
F_{\a_{n}\b_{n}}.
\end{equation}
This current is not divergenceless, because of  the sign function,
but it is cancelled by contribution
from the fermion zero modes trapped on the wall\cite{9}.
This is in accord with the general expectation that there is no
anomaly of continuous symmetry in odd dimensions.

Superficially it appears that one has successfully devised a
2n dimensional chiral gauge theory, or
at least a model of chiral fermions
coupled to gauge fields. One must, as Kaplan pointed out,
be careful about the gauge  field dynamics, making sure it is correct in
2n dimensional sense.
Kaplan argues that for a continuum Yang-Mills action
\begin{equation}
L_{YM} =\b_{1} F_{ij}^{2} + \beta_{2}F_{is}^{2},
\end{equation}
the equation of motion for $A_{\m}$ is
\begin{eqnarray}
\b_{1} D_{j}F_{ij} + D_{s}\b_{2}F_{is} &=&  J_{i}, \nonumber \\
\b_{2}D_{i}F_{si} &=& J_{s}.
\end{eqnarray}
In general $J_{s}$ does not vanish, so one cannot set $\b_{2}$ to zero.
One is not studying 2n dimensional gauge theory.

However for the spacial case of n=1, the long wave length of the gauge
fields off the wall can be derived  effectively from the following action
\begin{equation}
L =\b_{1} F_{ij}^{2} + \beta_{2}F_{is}^{2} + {\frac{{\rm sgn} m(s)}
{4\pi}} (AdA + {\frac{2}{3}}A^{3}).
\end{equation}
This is more or less the same as the topologically massive gauge theory
\cite{99,6}, except for some position dependent coupling constants.
It simply implies that there is nothing in low-energy physics
in 2+1 dimensions, since the gauge bosons are massive. More correctly,
it is a Chern-Simons topological field theory, which is
known to possess no local propagating degrees of freedom \cite{10}.
The scale at which nontrivial dynamics sets in is given by  $1/\b_{2}$.
On the other hand the typical energy scale in 2 dimensions is
$\sqrt{m/\b_{2}}$. The latter can be much lower than the former if
$m << \b$.
So a  generically non-zero
$\beta_{2}$ will still have no effect on the low-energy
dynamics if $\beta_{2}(0)=0$ and otherwise $\beta_{1}$,
and $\b_{1}$ is tuned to be much smaller than
$1/m$. In this particular
case, the model does look like a 2 dimensional theory. The theory
could be termed ``anomalous", because the chiral fermions  do produce
anomaly. But it is cancelled because of  the massive fermion
contribution from the extra dimension. For $n \ge 2$, the resulting
Chern-Simons term has no low energy effect, so that the gauge bosons
are massless in $2n+1$ dimensions, and the regularization fails to
produce a truly $2n$ dimensional theory. Note even if the gauge anomaly
cancels, one must still tune the parameters to cancel induced
$F_{si}^{2}$ term from matter fields.

We put quotation marks on the word anomalous because, as is well-known,
a 2D  ``anomalous" gauge theory is not really anomalous, because the gauge
degrees of freedom happen to be described by a chiral Lagrangian as well,
and the resulting model is renormalizable and can even be finite.
In our case, the extra degrees of freedom are obviously related to
the  Chern-Simons gauge theory with a boundary \cite{10,11},
which is a chiral WZW theory by itself.  In fact, the gapless
edge excitation of a quantum Hall system is a physical realization of
2D chiral gauge theory \cite{12}.

In conclusion, we have shown that Kaplan's regularization works for
``anomalous" gauge theories in D=2 as well.
Our observation further  supports the viability of that proposal.

\vspace{0.3in}
\begin{flushleft}
{\Large Acknowledgement}
\end{flushleft}
\support

\end{document}